# Tuning the conductance of single walled carbon nanotubes by ion irradiation in the Anderson localization regime

C. Gómez-Navarro[1], P.J. de Pablo[1], J.Gómez-Herrero[1],
B.Biel[2], F.J.Garcia-Vidal[2], A.Rubio[3], and F. Flores[2]

[1] Departamento de Física de la Materia Condensada, Universidad Autónoma de Madrid, E-28049 Madrid, Spain
[2] Departamento de Física Teórica de la Materia Condensada, Universidad Autónoma de Madrid, E-28049 Madrid, Spain
[3] Departamento de Física de Materiales, Universidad del País Vasco UPV/EHU and Donostia International Physics Center (DIPC), E-20018-San Sebastián, Spain

Carbon nanotubes[1,2] are a good realization of one-dimensional crystals where basic science and potential nanodevice applications merge[3]. Defects are known to modify the electrical resistance of carbon nanotubes[4]. They can be present in as-grown carbon nanotubes, but controlling externally their density opens a path towards the tuning of the nanotube electronic characteristics. In this work consecutive $Ar^+$ irradiation doses are applied to single-walled nanotubes (SWNTs) producing a uniform density of defects. After each dose, the room temperature resistance versus SWNT-length [$R(L)$] along the nanotube is measured. Our data show an exponential dependence of $R(L)$ indicating that the system is within the strong Anderson localization regime. Theoretical simulations demonstrate that mainly di-vacancies contribute to the resistance increase induced by irradiation and that just a 0.03% of di-vacancies produces an increase of three orders of magnitude in the resistance of a 400 nm SWNT length.

The traditional approximation to reduce the size and enhance the performance of electronic devices may not be applicable in the near future[5]. Electronic circuits based on molecules have created great expectation for their new foresighted properties. For the case of electronic circuits based on carbon nanotubes[6], the influence of disorder and defects[4,7] is of fundamental relevance in the performance of the device. In particular, the density of defects would determine the transport in nanotubes from a ballistic regime[8,9] to either weak or strong localization regimes. Quantum theory dictates that for a one dimensional conductor of length $L$[10,11], with a given density of defects, localization effects emerge when the "phase coherence length" $L_\phi$ is larger than the localization length $L_0$. If $L$ is not too large (for $L$ about 3-10 $L_0$) and the inelastic interaction is weak, the wire resistance is controlled by the phase-coherent electron propagation[12], falling into the strong localization regime in which the resistance increases exponentially with the length of the wire. This regime has not been observed in single-walled nanotubes in spite of the many evidences for weak localization diffusive regime and quantum interference in multi-walled carbon nanotubes[13]. By changing the density of defects, $L_0$ can be modified allowing to control the resistance of the one dimensional conductor.

Induced defects have been already used to modify different properties of carbon nanotubes. Indeed, electron-beam has been used to create in-situ nanotube junctions[14] and to enhance the mechanical response of nanotubes bundles by creating stable links among the tubes[15]. Theoretically, the effects of vacancies[16], topological defects[17] and random disorder[18,19] on the nanotubes electronic transport have been investigated. Surprisingly, a reduced number of experimental results have been published concerning the transport regime of localization and, to our knowledge, no experimental evidence of the influence of controllably induced defects on the variation of the electrical resistance with the length of single-walled carbon nanotubes has been reported yet[20]. In order to study this issue we irradiate SWNTs with $Ar^+$ ions and measure the electrical characteristics *of the same metallic carbon nanotube,* after each irradiation, by using a conductive AFM (Nanotec Electrónica S.L.) at room temperature. In addition, the measured low voltage resistance (LVR) data is theoretically analysed using a combination of density functional methods and Green's functions techniques.

A metallic AFM tip was used to measure the current *vs.* voltage characteristics (IV) of the nanotubes as a function of the distance between the metallic AFM tip, used as mobile electrode, and a fixed macroscopic gold electrode[21]. Briefly, the experiments were performed as follows: a metallic SWNT was first selected (Fig. 1a) and electrically characterized. From the IVs the LVR was determined along the tube by fitting a straight line to the data around zero volts. In this way a resistance *vs.* length curve *R(L)* is obtained (Fig. 1a). The experimental set up enables us to increase the force applied by the mobile electrode to the SWNT until an optimum contact resistance is reached[22].

All the individual metallic SWNTs probed in our experiments exhibit a non-linear *R(L)* at low voltage as expected for a strong Anderson localization regime[10]. The low-bias resistance data can be fitted by:

$$R(L) = R_c + \frac{1}{2} R_0 \exp(L/L_0) \qquad (1)$$

where $R_c$ is the contact resistance, $R_0$ is the inverse of the quantum of conductance $G_0 = 2e^2/h$ (the *1/2* factor in equation (1) accounts for the two conductance channels of a metallic SWNT, see below) and $L_0$ is the localization length. The exponential resistance comes from an interference effect by coherent backscattering of electrons at the defects. Fig. 1b shows *R(L)* for the selected nanotube before ion irradiation. This 1.4 nm diameter SWNT exhibits metallic IV characteristics, compatible with a (10,10) armchair nanotube. The furthest IV curve was acquired at a distance of ~700 nm (the uncover length of the nanotube) from the gold electrode. The closest one is taken at the minimum distance for which the tip is clearly not in contact with the gold electrode i.e. 50 nm. $R_c = R(0)$ is linearly extrapolated from the closest electrode measurements to be 50 KΩ. This is a typical contact resistance for individual SWNTs using gold contacts[8]. By fitting the data to equation (1), a value of *$L_0$= 216 nm* is found. Repeating the measurements along the same nanotube we obtain the same $L_0$ within a 5% error, indicating that the AFM tip does not introduce any measurable damage along the SWNT. In the case of 14 different metallic SWNTs we find an average localization length *$L_0$=231* nm with a standard deviation of ±117 nm (see Fig. 1 for different R(L) set of data in three different nanotubes). We have examined nanotubes from two different sources, HipCo (Carbon nanotechnologies ) and arc-discharged (M.T. Martinez group at Instituto de Carboquimica, Zaragoza, Spain) and both presented similar features, indicating that defect control in SWNT synthesis is still an issue. High-quality metallic carbon nanotubes grown by carbon vapour deposition exhibit very long low-bias mean-free-path as large as one micron, the resistance being controlled by the weak scattering with acoustic phonons[8,9]. However, as the scope of the present work is to assess the dependence of the localization length in metallic carbon nanotubes on defect concentration and tube-length, the quality of the initial nanotubes sample is not relevant.

After this first characterization, the nanotube sample was irradiated with an Ar$^+$ ion beam with energy of 120 eV (see methods section for details). After each ion dose the same nanotube is again electrically characterized finding a decrease in the localization length from 216 nm, before irradiation, to 61 nm for the 4$^{th}$ dose (the data are summarised in Fig. 2a). Notice that this drop in the localization length causes a change in resistance for L=400nm of about two orders of magnitude. Further irradiation doses (not shown) produced even higher resistances at this point. Taking into account the ionic current, the sample area, the irradiation time, the energy and assuming a (10,10) nanotube, the number of Ar$^+$ ions colliding per SWNT nanometer length can be estimated (see table in Fig. 4c). This estimation is corrected by a factor 1.3 when the trajectories of the Ar$^+$ ions for the experimental electrode geometry are simulated (SIMION 3D 7.0). In addition to the increment on the LVR of the SWNT, Fig 2a also shows a small increment of the contact resistance with the irradiation time that we attribute to the sputtering of gold and carbon atoms along the nanotube-electrode contact region.

Using molecular dynamic simulations, Krasheninnikov et al.[23] suggested that Ar$^+$ collisions mainly create mono-vacancies and di-vacancies along the nanotube.

In particular, their simulations for a 120 eV Ar$^+$ ion beam impinging perpendicular to the tube axis of a (10,10) nanotube show that di-vacancies appear in about 30-40% of the impacts (private communication). This is important for the rational behind the observation of the strong localization regime and in quantitative agreement with the estimations from our experiments, as discussed below.

In this work we have performed simulations using a first-principles Local Orbital Density Functional method[24] to calculate the relaxation around the defects and the effective Local Orbital Hamiltonian (a sp$^3$-basis set of FIREBALL orbitals was used with a cut-off radius of 2.15Å ). The advantage of this approach is that it provides a means of calculating the LVR using standard Green-function techniques derived for tight-binding Hamiltonians[25] but now with first-principles accuracy. Within this theoretical framework, as explained in the methods section, we are able to calculate the differential conductance $g$ for a realistic tube with a given number of defects (mono-vacancies or di-vacancies) that are randomly distributed along the nanotube (Fig. 3a). This numerical tool allows us to make a quantitative comparison between theory and experiments.

Before discussing the behaviour of long tubes it is worth discussing the low-bias conductance drop associated with a single defect (either mono-vacancy or di-vacancy) at zero temperature. A perfect (10,10) metallic SWNT presents a conductance of 2$G_0$. Our calculations show a conductance of 1.98 $G_0$ for the mono-vacancy and 1.36 $G_0$ for the *lateral* di-vacancy (see inset in Fig. 3b). Clearly, the mono-vacancy represents a very small scattering to the electron propagation along the nanotube compared with the di-vacancy. This already suggests that we can safely neglect the mono-vacancy effect on the final low-bias conductance and concentrate on analysing the di-vacancies (this is confirmed by extensive calculations including mono- and di-vacancies that are not presented here).

Fig. 3b portrays the nanotube LVR at zero temperature calculated as a function of its length for a mean distance between di-vacancies, $d$=28.5 nm (the length of the nanotube can be inferred by just multiplying the number of defects by $d$). Several cases are shown corresponding to different random configurations of di-vacancies; the LVR mean value in the log scale[26] is also presented (black line), and was obtained after averaging over 90 random cases (the grey region in Fig. 3b represents the mean square root deviation). Remarkably, an exponential dependence of the LVR mean value versus length of the nanotubes is clearly present in our simulations. This result is general and supports the view that, in the experiments, Anderson localization is causing the observed increase in resistance. However, the strong conductance fluctuations calculated for each configuration (Fig. 3b) are not seen in our experimental data. This is due to the energy window of the injected electrons, of the order of $k_BT$, that introduces significant differences in their phase-coherent propagation along the tube. If $L_\phi$ is much larger than $L_0$ (as it is the case at low bias in our samples), the condition $L_T \leq L_0$, (with $L_T = \dfrac{\hbar v_F}{k_B T}$ being the thermal coherent length and $v_F$ the Fermi velocity), defines the regime for which the phase shifts of the electrons (and their corresponding conductances) when travelling through $L_0$ change significantly within the energy window. Then, the conductance of each and every single electron does indeed fluctuate with the nanotube length but, importantly, these fluctuations are different for electrons injected with different energies. Therefore, it is expected that the measured conductance, that has the contribution of a large number of electrons, would not fluctuate yet the exponential decay in the conductance being still present. This line of reasoning is corroborated by our numerical simulations. In Fig. 3c we illustrate these thermal effects by showing our calculated LVR at room temperature for two of the random configurations of defects analyzed in Fig.3b. Clearly the LVR now presents a smooth exponential behaviour that is similar (within a 30 % accuracy) to the mean value found for the random configurations. The low-bias conductance $G$ (LVR is just the inverse of $G$) at room temperature is calculated by integrating the differential conductance, $g$, over an energy window defined by the derivative of the Fermi distribution function (see methods section for details). The inset in Fig.3c shows $g$ as a function of electron energy for one of the random configurations containing 25 defects. The sharp peaks and the strong fluctuations of that conductance are a clear indication of the strong Anderson localization regime for the electrons propagating along the nanotube. For zero temperature, $G=g(E_F)$ and then $G$ is extremely sensitive to the particular configuration of defects. However, at room temperature, the integration over the energy window eliminates the fluctuations in the conductance (or resistance) as a function of the nanotube length, still yielding an exponential behaviour in the LVR, in agreement with the qualitative arguments presented above. In this paper, we have calculated the LVR using only the mean value of the conductance for the random configurations at the Fermi level, an approximation that yields a reasonable accuracy to the LVR at room temperature, although an improved analysis should incorporate thermal effects for all the nanotube lengths and all the random configurations (work along this line is in progress in our lab).

Fig. 4a shows the averaged LVR calculated for different density of defects in the nanotubes with $d$ ranging from 2.9 to 75.5 nm. In all cases we observe the strong localization regime. To make more evident this localization effect we show in Fig. 4b the calculated values for the localization length, $L_0$, as a function of $d$. We get two distinct regimes: i) for lower defect density (i.e., large values of $d$, $d>5$ nm), $L_0$ depends linearly with $d$, $L_0 \approx 4.1 \cdot d$. ii) For higher defect density (small values of $d$),

$L_0$ saturates to a value close to 23 nm. In our experiments the defect density is always below 0.15% corresponding to case (i).

In order to compare quantitatively experiments and theory we have to take into account the initial density of defects (most likely substitutional[27]) of the nanotube before irradiation giving rise to the initial localization length $(L_0)_{in}$ of 216 nm. Assuming that the total localization length is obtained via Mathiessen's rule as $[(L_0)_{total}]^{-1} = [(L_0)_{defect}]^{-1} + [(L_0)_{in}]^{-1}$ ( $(L_0)_{defect}$ being the localization length due to defects induced upon irradiation), $(L_0)_{defect}$ is easily found for each irradiation dose from the experimental $(L_0)_{total}$ obtained using (1). Once $(L_0)_{defect}$ is known from the experiments it can be compared with the calculated $L_0$. This allows us estimating the average distance between di-vacancies, d, for each $L_0$. If we now compare $d$ with the estimated number of ion collisions per nm we infer an efficiency of the $Ar^+$ ion beam of one di-vacancy every four ion collisions, which is in quantitative agreement with Krasheninnikov et al.[23]. The experimental and theoretical results are summarised in the table shown in Fig. 4c.

These arguments combined with the data presented in Fig. 4 tell us that, due to localization effects, only 0.03% of di-vacancies produce an increment of three orders of magnitude in the resistance of a 400 nm length segment of a metallic nanotube. This explains why, although the consequences of the presence of defects in the nanotube electronic transport are so noticeable, we have not observed any morphological change within the resolution of the AFM in the nanotube structure under irradiation, even when the tube electrically fails at *L = 420 nm* in the last irradiation dose (see Fig. 2 and table in Fig. 4c).

For the sake of completeness it is worth discussing briefly the high voltage case. As our experimental IVs were obtained by ramping the voltage from -2.5 eV up to +2.5 eV, the high voltage resistance (HVR) can also be obtained, allowing us to assess the influence of defects in the HVR. This study is of relevance for nanotube applications as the current carrying capacity of perfect carbon nanotubes is limited by phonon emission. In fact, recent measurements [9,12] in high-quality metallic nanotubes show a very short electron mean free path, 10-30 nm at high bias due to scattering with optical phonons [28]. From the point of view of the electron propagating along the wire, the main effect of a high voltage is to substantially increase its electron-phonon inelastic scattering which tends to favour a diffusive conductivity in the sample. This is seen in our experiments for voltages higher than 0.3 eV: in this limit, the nanowire resistance shows a linear dependence versus length, instead of the exponential behaviour found at low bias, confirming the change in the conductance regime. A set of measured IVs (see Fig. 2b) acquired at the same spot of a metallic SWNT for different irradiation doses exhibit a LVR≈ 0.3·HVR for the SWNT before ion irradiation; as the number of defects grows the situation is reversed and eventually the experimental data show LVR≈ 3·HVR. In this case the HVR is mainly dictated by electron-phonon coupling whereas the LVR exhibits strong signatures of Anderson localization induced by the defects. By taking the measured HVR as a function of the length for each irradiation dose, we can estimate the mean free path $\ell$, as $\ell = R_0 L/R$ *(9)*. This shows that $\ell$ varies from 30 nm before irradiation, in good agreement with electron mean free paths due to the electron-phonon interaction [9,12], to 12 nm for the highest irradiation dose. A careful theoretical analysis is required to better understand these results.

In conclusion, we have shown the extreme importance of defects (in particular di-vacancies) on the low-bias conducting properties of single-walled carbon nanotubes irradiated with an $Ar^+$ ion beam. We have shown that only a 0.03% of di-vacancies produce an increment of three orders of magnitude in the resistance of a 400 nm carbon nanotube segment. Numerical simulations indicated that the irradiation efficiency is of one di-vacancy per every four ions collision. The experimental data revealed a room-temperature exponential dependence of the low-bias resistance versus nanotubes length, as expected for a strong Anderson localization regime. Our extensive theoretical calculations support this conclusion and illustrate how the short thermal length at room temperature cancels the oscillations on the LVR(L). By comparing the low and high bias resistance we also observed a cross-over between a defect-controlled transport (low-bias) to phonon-mediated one (at high-bias). Besides its fundamental relevance, this work opens new paths to tailor the electrical properties of future nanotube devices using ion irradiation.

## Methods

### Experiments

The irradiation is performed at an Ar pressure of $1.0 \times 10^{-4}$ mbar. The ion energy used is 120 eV. We apply irradiation doses during consecutive periods of 5 or 10 seconds. A 0.1 µA ionic current is measured on the macroscopic electrode (1.8 cm$^2$ area) where the nanotubes are contacted.

Below 70 eV energy we do not observe any conductance drop in the irradiated SWNTs. For doses longer than one minute at 120 eV the resistance of the irradiated sample is so high that the current flowing through the individual SWNTs is below our sensitivity (~0.1 nA).

### Theory

The simulation geometry consists of a device region with the defected nanotubes connected to two semi-infinite perfect tubes ($S_1$ and $S_2$) that act as left and right electrodes (see Fig.3a). In this approach, the differential conductance g(E) is calculated at the tube-lead interface $S_2$ from:

$$g(E) = \frac{4\pi e^2}{\hbar} Tr\left[\hat{D}_{LL}^{A}(E)\hat{T}_{L2}\hat{\rho}_{22}(E)\hat{T}_{2L}\hat{D}_{LL}^{R}(E)\hat{\rho}_{LL}(E)\right] \quad (2)$$

where $Tr$ represents the trace of the operator in brackets. $\hat{T}_{L2}$ describes the coupling between the right end of the nanotube and lead 2, $\hat{\rho}_{22}(E)$ being the density of states matrix associated with the decoupled ($\hat{T}_{L2} = 0$) semi-infinite lead 2 projected at $S_2$ and related to the retarded Green's function ($\hat{g}_{22}^{R}$) by $\hat{\rho}_{22} = -\frac{1}{\pi} \text{Im}\, \hat{g}_{22}^{R}$. For calculating $\hat{g}_{22}^{R}$ we use standard decimation techniques. $\hat{\rho}_{LL}(E)$ is the density of states matrix projected onto the right end of the nanotube and is calculated by an iterative procedure which is started on the side $S_1$ and includes, at each step, a new layer of the tube. After $N$ steps, we build the whole tube and their corresponding retarded and advanced Green's functions $\hat{g}_{LL}^{R(A)}$ (with the projection of the electronic states of lead 1 incorporated). In this way, we can include straightforwardly a random configuration of defects distributed along the nanotube with a given mean distance between them (see Fig.3a). The retarded and advanced denominator functions $\hat{D}$ appearing in equation (2) are given by:

$$\hat{D}_{LL}^{R(A)}(E) = \left[I - \hat{g}_{LL}^{R(A)}(E)\hat{T}_{L2}\hat{g}_{22}^{R(A)}(E)\hat{T}_{2L}\right]^{-1} \quad (3)$$

At zero-temperature, the low bias conductance G is calculated by just evaluating $g$ at the Fermi energy, $g(E_F)$. If we are interested in obtaining G at a temperature T, we have to evaluate the following integral: $G = \int_{-\infty}^{\infty}(-\frac{df_T}{dE})g(E)dE$ where $f_T(E) = \frac{1}{e^{\frac{E-E_F}{k_B T}}+1}$ is the Fermi distribution function and $k_B$ the Boltzmann constant.

Within this theoretical framework, it is feasible to calculate the low bias conductance of very long nanotubes (up to several microns long) with an arbitrary distribution of defects immersed on them.

**Acknowledgments:**
We thank Sidney Davison and Ron Reifenberger for careful reading and Jose Ortega for helping with numerical methods. This work was partially supported by Spanish MCyT under contracts MAT2001-00664, MAT2001-00665 and MAT2002-01534 and the European Community IST-2001-38052 and NMP4-CT-2004-500198 grants.

Correspondence and requests for materials should be addressed to J. G. e-mail: julio.gomez@uam.es

# Figure captions

**Figure 1. Experimental set up.**

Fig. 1a) AFM image (1µm x 1µm) of a SWNT adsorbed on an insulating substrate connected to a gold electrode (bottom). The inset is a scheme of the experimental set-up showing a gold covered AFM tip, the macroscopic gold electrode, the SWNT and the used circuit.

b) c) d) Plots of the low-voltage-resistance *vs*. length for three metallic SWNTs as deposited on the surface, without irradiation. The data are fitted to equation (1). The values of $R_c$ and $L_0$ obtained for the best fit are depicted in each char.

For Fig. 1b), data for this nanotube after irradiation are presented in Fig. 2.

**Figure 2. Effect of consecutive irradiations on the electrical resistance.**

Fig. 2a) Semi-logarithmic plot of the LVR vs. L data for a 1.4 nm SWNT diameter before and after each of the four consecutive irradiation doses, corresponding to the $1^{st}$, $2^{nd}$, $3^{rd}$ and $4^{th}$ irradiations.

b) Differential conductance (dI/dV) vs. bias voltage. The curves were acquired at a distance of 350 nm between electrodes. The cross-over between LVR and HVR is clearly seen.

**Figure 3. Simulated resistance *vs*. length for a fixed density of defects**

There are two possible orientations for the di-vacancies. We define the orientation of the di-vacancy as the vector connecting the two missing atoms. The *vertical* case (with its orientation perpendicular to the nanotube direction) is less stable than the *lateral* one (with their orientation partially parallel to the tube axis) by 2.0 eV; then we can safely assume that $Ar^+$-irradiation only creates either mono- or lateral di-vacancies. Inset in Fig. 3b displays the relaxed geometry of a lateral di-vacancy in a (10,10) SWNT. Fig. 3b shows the nanotube resistance as a function of its length for a mean distance between di-vacancies of 28.5 nm. Several cases are illustrated corresponding to different random configurations of the di-vacancies (see Fig. 3a); the conductance mean value (thick black line) and the dispersion (dotted lines) are also shown in Fig. 3b, being obtained after averaging over 90 random cases. In this figure, the resistance is shown just after each di-vacancy plane, and the length of the nanotube is obtained by multiplying the number of defects in the x-axis by the mean distance between the di-vacancies. A summary of the results for other defect concentrations is provided in Fig. 4. Panel c shows the nanotube resistance for two of the random configurations analyzed in Fig.3b evaluated at zero temperature (dashed lines) and at room temperature (solid lines). The inset shows the differential conductance (in logarithm scale) for the configuration displayed in red and for a defected nanotube of length 712 nm (25 defects), only in the range $(-k_BT, +k_BT)$.

**Figure 4. Localization length: experimental and theoretical results.**
Fig. 4 a) Plot of the calculated mean value of *ln(2R/R$_0$)* as a function of the number of defects, for different *d´s*. At long distances, *R* can be fitted to *R$_0$ /2 exp(L/L$_0$)*, *L$_0$* being the localization length. Fig. 4b shows our calculated values of *L$_0$* as a function of *d*: for large values of *d* (*d >5* nm), *L$_0$* approaches *4.1·d*; for small values of *d*, *L$_0$* saturates to around 23 nm. c) Summary table showing the relevant figures of experiment and theory. The shadowed columns are directly obtained from the experiments. The values in the sixth column have been obtained via Mathiessen's rule from the experimental *(L$_0$)$_{total}$*. The seventh column are the data corresponding to the average distance between di-vacancies obtained theoretically for the values of *(L$_0$)$_{defect}$* in the sixth column.

Fig. 1

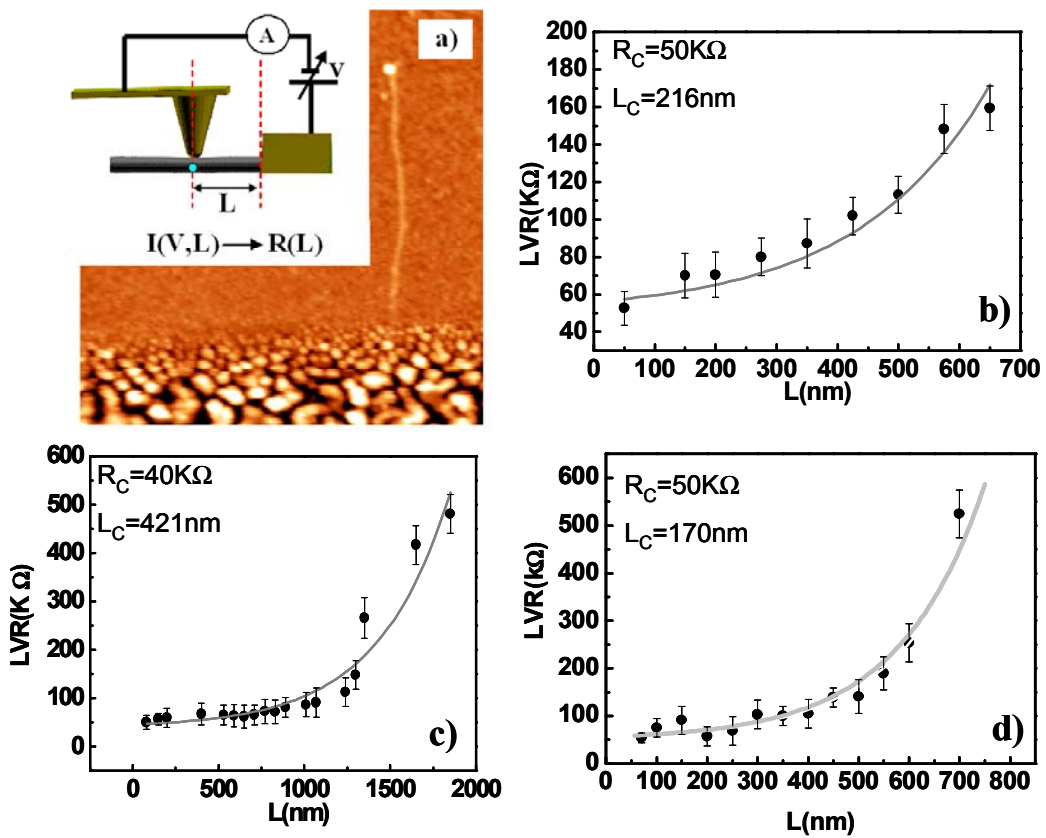

Fig. 2

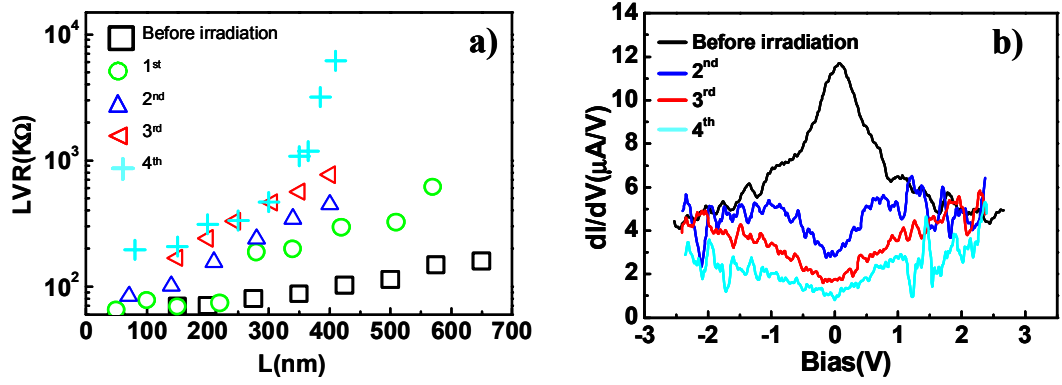

Fig. 3

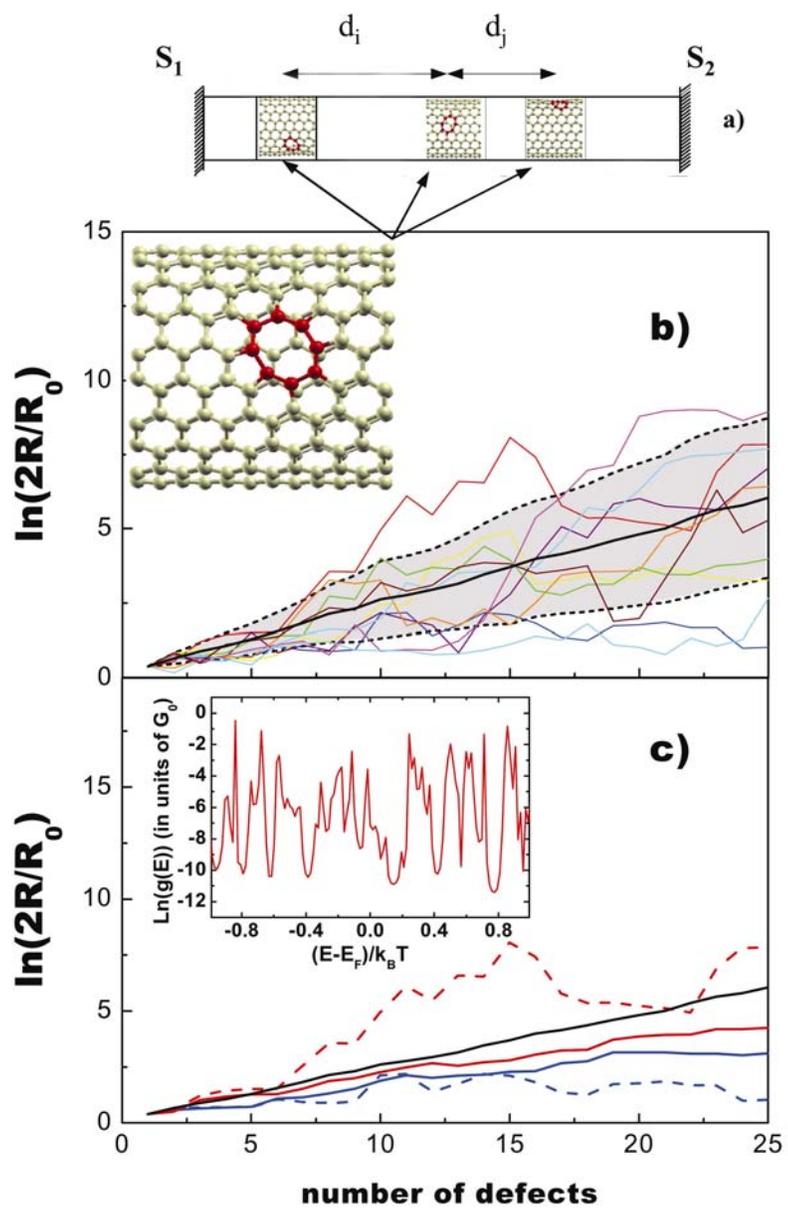

Fig. 4

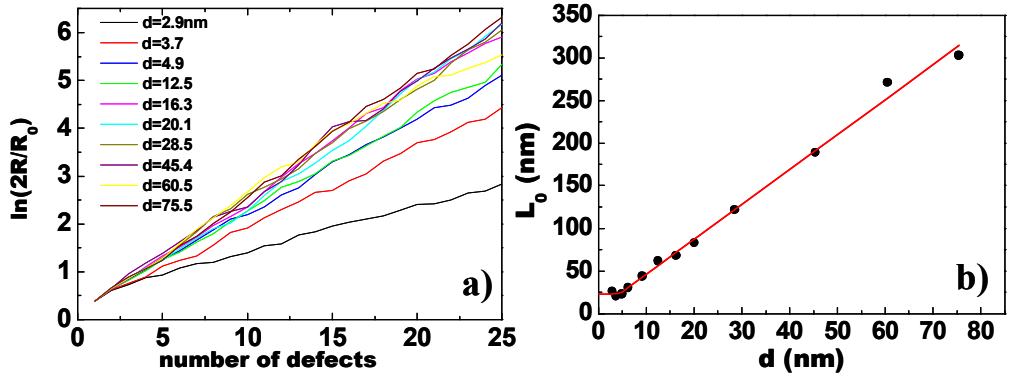

| c) Irradiation # | Irradiation time (s) | colliding Ar$^+$ ion /C atom (%) | Average distance between colliding ion $d_{ion}$ (nm) | $(L_0)_{total}$ (nm) | (*) $(L_0)_{def}$ (nm) | (**)Average distance between divacancies d(nm) |
|---|---|---|---|---|---|---|
| 0 | 0 | 0 |  | 215.9 |  |  |
| 1$^{st}$ | 10 | 0.043 | 14.4 | 126.9 | 308 | 75 |
| 2$^{nd}$ | 15 | 0.064 | 9.6 | 94.5 | 168 | 41 |
| 3$^{rd}$ | 20 | 0.086 | 7.2 | 83.7 | 137 | 33 |
| 4$^{th}$ | 25 | 0.107 | 5.75 | 61.1 | 85 | 21 |
| 5$^{th}$ | 30 | 0.129 | 4.8 | 45.2 | 57 | 14 |

(*) Estimated from the measured $(L_0)_{total}$ and $1/(L_0)_{ini}$ using $1/(L_0)_{total} = 1/(L_0)_{ini} + 1/(L_0)_{defect}$

(**) Obtained using the calculated relation $\langle d \rangle \approx (L_0)/4.1$ from panel b)